\documentstyle[12pt]{article}
\newcommand{\m}{\medbreak}

\newcommand{\no}{\noindent}
\newcommand{\EQ}{\begin{equation}}
\newcommand{\eq}{\end{equation}}
\newcommand{\EQA}{\begin{eqnarray}}
\newcommand{\eqa}{\end{eqnarray}}

\newcommand{\AR}{\renewcommand {\arraystretch}{1.5}
\begin{array}{l}}
\newcommand{\bAR}{\renewcommand {\arraystretch}{2}
\begin{array}{l}}
\newcommand{\ARc}{\renewcommand {\arraystretch}{1.5}
\begin{array}{c}}
\newcommand{\bARc}{\renewcommand {\arraystretch}{2}
\begin{array}{c}}
\newcommand{\ar}{\end{array} \renewcommand {\arraystretch}{1}}

\begin{document}
\begin{titlepage}
\vspace{0.2in}
\vspace*{1.5cm}
\begin{center}
{\large \bf
 Spin asymmetries from Non Standard Physics\\
with polarized proton beams at RHIC
\\}
\vspace*{0.8cm}
{\bf P. Taxil} and {\bf J.-M. Virey}{$^1$}  \\ \vspace*{1cm}
Centre de Physique Th\'eorique$^{\ast}$, C.N.R.S. - Luminy,
Case 907\\
F-13288 Marseille Cedex 9, France\\ \vspace*{0.2cm}
and \\ \vspace*{0.2cm}
Universit\'e de Provence, Marseille, France\\
\vspace*{1.8cm}
{\bf Abstract \\}
\end{center}
A new handed interaction between quarks could be at the origin
of some small parity violating effects in the inclusive production
of one-jet in polarized $pp$ collisions. This interaction could
originate from compositeness or from the presence of a new massive
vector boson. The  measurement of such asymmetries could be performed
within a few  years by using the RHIC collider as a polarized
proton-proton collider.

\begin{center}
{\it To appear in the proceedings of the\\
$12^{th}$ International Symposium on High Energy Spin Physics.}
\end{center}

\vfill
\begin{flushleft}
\m\no
Number of figures : 1\\

\m\no
October 1996\\
CPT-96/P.3391\\
\m\no
anonymous ftp or gopher : cpt.univ-mrs.fr

------------------------------------\\
$^{\ast}$Unit\'e Propre de Recherche 7061

{$^1$} Moniteur CIES and allocataire MESR \\
email : Taxil@cpt.univ-mrs.fr
\end{flushleft}
\end{titlepage}


In a few years from now, the RHIC Spin Collaboration (RSC)
will use the RHIC machine as a polarized $pp$ collider.
The center of mass energy will be as high as 500 GeV, the
luminosity ${\cal L} = 2. 10^{32} \, cm^{-2}.s^{-1}$ and the degree
of beam polarization around 70\% [1]. With these figures,
spin asymmetries as small as 1\% should be measurable in a few
months running. For example, the parity
conserving (PC) double-spin longitudinal asymmetry $A_{LL}$ in
inclusive one-jet production will be obtained with a very small error,
hence allowing to test the spin structure of QCD.
Adding the measurement of some parity violating (PV) asymmetries
in $W$ and $Z$ boson production, it will be possible to
determine very accurately the various polarized partonic
distributions inside the proton (see [1,2] and also e.g. [3]).
On the other hand, as was noticed some time ago [4],
non conventional PV effects in hard-hadron reactions have never
been searched for. Focusing on the inclusive production of one jet,
one can define a double  helicity PV asymmetry $A_{LL}^{PV}$ as:
\begin{equation}
A_{LL}^{PV}
={d\sigma_{(-)(-)}-d\sigma_{(+)(+)}
\over  d\sigma_{(-)(-)}+d\sigma_{(+)(+)}}
\end{equation}
\noindent
where $d\sigma_{\lambda,\lambda'}$ means the cross
section in a given helicity configuration for the production
of a single jet with transverse energy $E_T$,
integrated over some rapidity interval around $y=0$.

According to the Standard Model, this process
is essentially governed by QCD plus a small contribution
due to electroweak boson exchanges. The latter induces a small
$A_{LL}^{PV} $ which should  be visible at RHIC.
On the other hand, a new interaction between quarks could be at the
origin of some deviations from the expected $A_{LL}^{PV} $, provided that
it presents a particular chiral structure.

For instance, in the framework of compositeness,
an effective handed interaction due to
a contact term between quarks can be present [5] :
\begin{equation}
{\cal L}_{qqqq} = \epsilon \, {\pi\over {2 \Lambda_{qqqq}^2}}
\, \bar \Psi \gamma_\mu (1 - \eta \gamma_5) \Psi . \bar \Psi
\gamma^\mu (1 - \eta \gamma_5) \Psi
\end{equation}
\noindent
where $\Psi$ is a quark doublet, $\Lambda$ is the compositeness scale,
$\epsilon$ and $\eta$ taking the values $\pm 1$.
Furthermore, the presence of a new massive vector boson $Z'$, with
right-handed and left-handed couplings to each quark of flavour $i$ :
\begin{equation}
{\cal L}_{Z'} = {g\over 2 \cos \theta_W} Z'^{\mu}{\bar q}_i
\gamma_\mu[ C_{iL} (1 -
\gamma_5) \; +\; C_{iR} (1 + \gamma_5) ] q_i
\end{equation} \noindent
could generate new amplitudes and lead to similar effects on the
PV spin  asymmetry provided its couplings to quarks are enhanced.
Such "hadrophilic" or "leptophobic" neutral vector bosons $Z'$
have been recently considered from the phenomenological  point of
view [6], in order to explain some possible deviations from the
Standard Model predictions observed both at LEP and at CDF.

In two recent papers [7,8], we have studied the possibilities offered
by the polarized RHIC collider for disentangling these new effects from
the Standard spin asymmetry $A_{LL}^{PV}(SM)$ wich is mainly due to
the interference between the one-gluon exchange amplitude and
the $Z^0$ exchange amplitude. An illustration is given in Fig. 1.
%
%
%
One can see that the Standard asymmetry (plain curve) rises with the
transverse jet energy. This is simply due to the increasing importance
of quark quark scattering relatively to other terms involving real
gluons. The asymmetry is small but clearly visible at RHIC (the error
bars refer to the statistical error with an integrated luminosity
$L = 3.2 fb^{-1}$). The dot-dashed curves show the effect
of a contact term with a compositeness scale $\Lambda = 1.6$ TeV which
corresponds to the present CDF bound. We have also assumed that
parity were  maximally violated by this interaction. The effect
on $A_{LL}^{PV} $ is clearly visible : if the contact term
is indeed present this  measurement should allow to
get a unique information on its chirality  structure since
the sign of $A_{LL}^{PV} $
is sensitive to the sign  of the product $\epsilon.\eta$
($\epsilon = -1$ means constructive interference, see [5,7]).
From the absence of
deviations from the expected $A_{LL}^{PV} $ value,
it will be possible to place a
bound on $\Lambda$ around 4 TeV, which competes with the
Tevatron for an integrated luminosity  $L=100 fb^{-1}\,$
(with unpolarized beams).

Concerning the effect of a new hadrophilic $Z'$, we restrict
for illustration to the first model in [6] in
which the new current exhibits a particular chiral structure with
a dominance of the right-handed couplings on the left-handed
couplings (for $u$ quarks $C_{uR} \approx 2.C_{uL}$).
As a consequence, $A_{LL}^{PV} $ is
negative : the dotted curve corresponds to a best fit value for
the parameters of the model and the dashed curve to the maximal
effect (the mass of the $Z'$ is taken to be $M_{Z'}=1\,$TeV).
Finally, the "SU(5)" curve corresponds to a "flipped SU(5)" gauge
model inspired by superstring theories [9]. In this case, leptophobia
occurs naturally as well as maximal parity violation in the up-quark
sector. Leptophobia implies that the traditional $Z'$ searches at future
collider through the leptonic decay channel will be hopeless. Therefore,
the PV spin asymmetry in the jet channel should be the best observable
to pin down the presence of such an exotic object.

\vspace{0.5cm}

In conclusion, we have seen that some deviations from the expected
Standard PV asymmetry which could be induced by New Physics
beyond the Standard Model should be visible
at RHIC,  despite  the relatively low value of the c.m energy.
This is mainly allowed by the high luminosity and by the high degree
of polarization of the beams.
Therefore, the very precise measurements performed by the RSC
could contribute significantly to the analyses of some possible
manifestations of New Physics beyond the Standard Model.

\newpage

{\large \bf References}
\vspace{0.5cm}
{\small\begin{description}
\item{[1]}
G. Bunce et al.,
{\it Polarized protons at RHIC}, Particle World,
{\bf 3}, 1 (1992)
\item{[2]}
see  H. En'yo, N. Hayashi, S.F. Heppelman, Y. Makdisi,
contributions to this conference.
\item{[3]}
C. Bourrely, J. Ph. Guillet and J. Soffer, Nucl. Phys. {\bf B361}, 72 (1991)
\item{[4]}
M. Tannenbaum, in {\it Polarized Collider Workshop}, J. Collins,
S.F. Heppelmann and R.W. Robinett eds., AIP Conf. Proceedings
{\bf223}, AIP, New York, 1990, p. 201.
\item{[5]}
E. Eichten, K.Lane and M. Peskin, Phys. Rev. Lett.
{\bf 50}, 811 (1983) ;\\ E. Eichten et al., Rev. Mod. Phys.
{\bf 56} (1984) 579
\item{[6]} G. Altarelli et al., Phys. Lett. {\bf B375}, 292 (1996);\\
P. Chiappetta et al. Phys. Rev. {\bf D54}, 789 (1996)
\item{[7]} P. Taxil and J.M. Virey, Phys. Lett. {\bf B364}, 181 (1995).
\item{[8]} P. Taxil and J.M. Virey, Phys. Lett. {\bf B383}, 355 (1996).
\item{[9]} J.L. Lopez and D.V. Nanopoulos [hep-ph 9605359]
\item{[10]} M. Gluck, E. Reya  and W. Vogelsang, Phys. Lett. {\bf B359},
 201 (1995)
\end{description}}

\vspace*{4cm}
{\bf Figure captions}
\bigbreak
\noindent
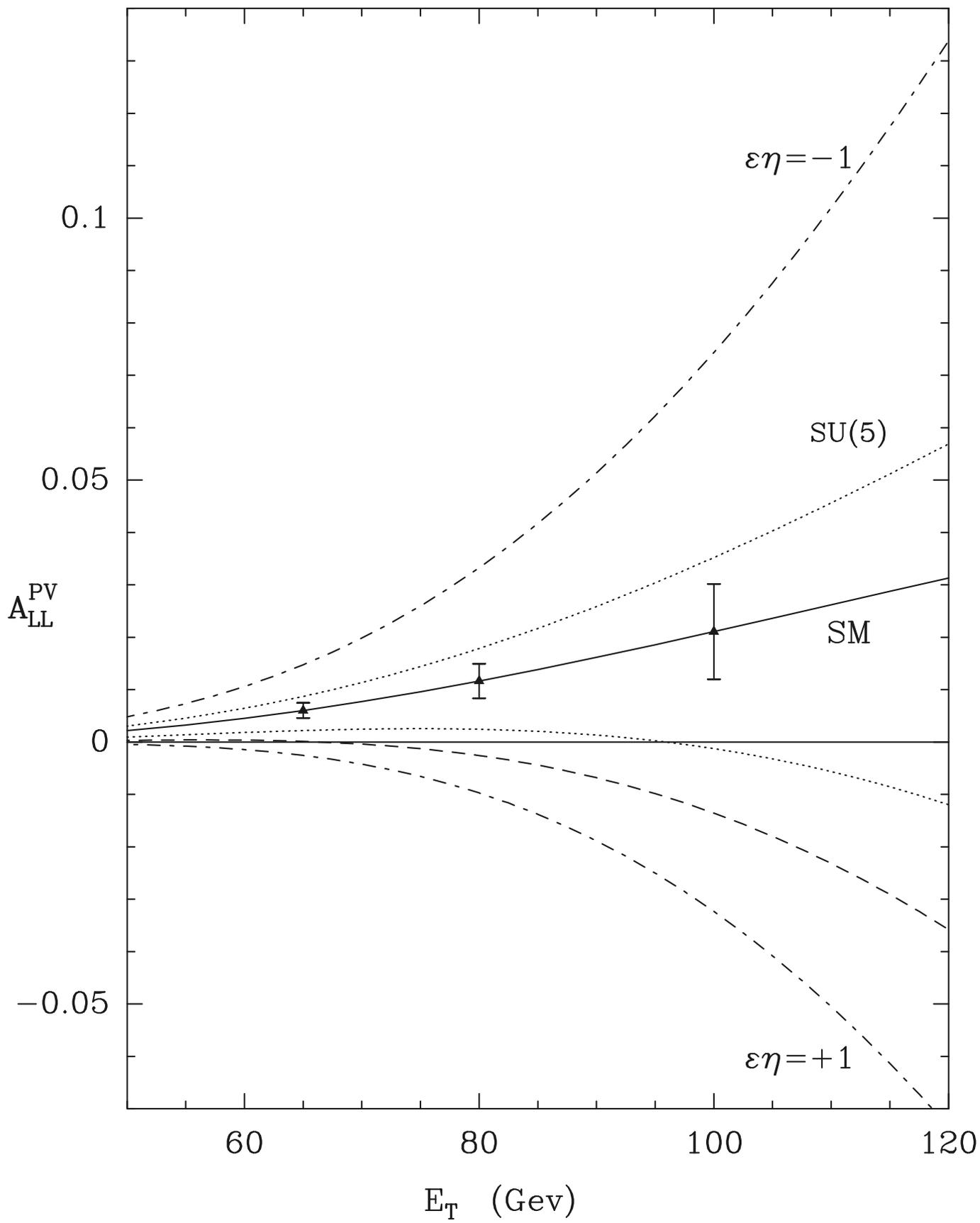
{\bf Fig. 1} The asymmetry $A_{LL}^{PV}$ for inclusive one-jet
production at RHIC as a function of the transverse jet energy $E_T$.
The calculations are performed with GRV [10] distributions.
\bigbreak
\noindent

\end{document}